\documentstyle[11pt,proceedings,twoside,epsf]{article}
\begin{document}

\title{The Arecibo Dual-Beam Survey: \\
Filling the Gaps in the Extragalactic Census}
\author{Jessica L. Rosenberg \& Stephen E. Schneider}
\affil{University of Massachusetts, Department of Astronomy, Lederle
Graduate Tower, Amherst, MA 01003}

\begin{abstract}

We present the Arecibo Dual-Beam Survey (ADBS), a ``blind" 21cm survey for
galaxies. The ADBS is the largest to date with the Arecibo radio telescope. We
were able to achieve an rms sensitivity of 3-4 mJy in only 7 seconds. The ADBS
covers substantially more volume than the Spitzak \& Schneider (1998) and the
Zwaan et al. (1997) surveys. It is surpassed by the Parkes HIPASS survey
(Kilborn et al. 1999) but has much better angular resolution and sensitivity. We
will discuss the HI sample and our efforts to quantify the completeness and
reliability. 

This is the first survey to use synthetic sources to establish the relation
between the flux, line width, and completeness of the survey. An empirical
understanding of these quantities is vital to the determination of the mass
function. Even if we do not understand all of the reasons why we do not detect
galaxies, the empirical understanding allows us to account for them properly.

In addition to the more general discussion of the ADBS survey, we will
discuss galaxies which have large ratios of HI mass to optical area and probably
have extended HI disks. We detect a much higher percentage of these high HI mass
to optical area galaxies than are seen in optical surveys indicating that
optical surveys miss these galaxies. Among these galaxies are also 7 galaxies
with HI masses $<$  10$^8$ M$_{\sun}$. Based on the mass function of Schneider
\& Rosenberg (this volume), the low mass galaxies are  an important part of the
general galaxy population.

\end{abstract}

\section{Introduction}

We present the Arecibo Dual-Beam Survey (ADBS) and describe the original 
driftscan detection survey, the VLA and Arecibo follow-up observations, 
and the use of synthetic sources as a way of evaluating sample completeness. 
We then discuss the indications from the ADBS sample that even low surface 
brightness optical surveys have overlooked a population of compact and low
surface brightness (LSB) galaxies that are found in HI surveys. The ADBS, and
other ``blind" HI surveys provide an important view of galaxy populations that
is completely unbiased by star formation. Comparing our HI-selected sample to
the dwarf and LSB sample of Schneider et al (1990, 1992), we find that the dwarf
and LSB sample misses a significant fraction of the LSB population. The
detection and study of this galaxy population is important for our understanding
of galaxy formation and evolution.

\section{The Arecibo Dual-Beam Survey}

\par The goals of our survey are (1) to determine whether there are
classes of galaxies that have been overlooked previously and (2) to help tie
down the HI mass function with the larger number of sources in this sample. To
achieve these goals a ``blind" HI survey needs to detect galaxies down to the
lowest possible HI fluxes. At least equally important is a detailed 
understanding of the survey limitations. The statistics of previous surveys 
have been insufficient to definitively establish the shape of the HI mass 
function at both the high and the low mass ends. 

Zwaan et al. (1997) and Spitzak \& Schneider (1998) undertook ``blind" HI 
surveys, similar to the one discussed here, but neither had enough low mass 
sources to make a strong claim about the shape of the mass function. Schneider, 
Spitzak, \& Rosenberg (1998) combined the two data sets and found evidence for a
rise in the number of low mass sources, but the results are not definitive. The
HIPASS survey, with its coverage of the entire Southern sky, should also
contribute to our understanding of the HI mass function. The most recently
published mass function contains 263 galaxies, comparable to the number in this
survey (Kilborn et al. 1999) but only has 2 galaxies with M$_{HI} < $ 10$^8$
M\sun\ (converting to H$_0$ = 75 km s$^{-1}$ Mpc$^{-1}$). HIPASS will cover
significantly more volume than the ADBS in the end, but there are still issues
that will make the detection of low mass  sources more difficult (1) the rms =
13 mJy which means that low mass sources are only detected very nearby where
determining the distance is difficult due to the effects of peculiar velocities
(2) the telescope beam is so large that source confusion is a problem. For
confused sources it has to be  determined, after the fact, whether any of them
could have been detected  independently.  We detect 7 sources with HI masses $<$
10$^8$ M\sun, almost as many as all of the previous ``blind" HI surveys
combined.

The Arecibo Dual-Beam Survey (ADBS) consisted of approximately 500 hours of
driftscan observations at the Arecibo 305 m telescope before the Gregorian
upgrade, between December 1993 and February 1994. These data consist of
$\sim$300,000 spectra taken every 7 seconds and cover nearly 24 hours of time in
each of 30 declination strips. The total sky coverage is approximately 430
deg$^2$ in the 3.3\arcmin\ main beam. In these data we find 265 galaxies that
were confirmed on follow-up.

Observing in driftscan mode meant that two feeds could be used simultaneously. 
We used the 21 cm and 22 cm circular polarization feeds, which have now been 
replaced by the Gregorian reflector. These were located on opposite
corners of the carriage house and pointed 1.6\deg\ apart on the sky. Our
ability to use two feeds doubled the area surveyed. The velocity coverage for 
the survey was --654 to 7977 km s$^{-1}$. This coverage was achieved over 512 
channels in each of two polarizations for each of two feeds, totalling 2048 
channels. The channel spacing was 16.9 km s$^{-1}$\ and the resolution was 
33.8 km s$^{-1}$\ after Hanning smoothing. The average rms noise for the 
survey was 3.5 mJy.

\par One of the major challenges of any single-dish HI survey is to distinguish
between real sources and radio interference. We used three tactics to address
this problem while also increasing the volume coverage and improving the
signal-to-noise: (1) Comparing data from the 21 and 22 cm feeds taken
simultaneously -- the interference usually enters through far sidelobes and
appears in both feeds while real sources appear in only one. (2) Comparing left
and right circular polarizations in each feed -- the average of the two
polarizations provides improved signal-to-noise while the difference is an
additional check for interference since it is often highly polarized. (3)
Observing each declination strip twice, on separate days, providing confirmation
of the source detections.

\par While the driftscan technique allowed us to cover a large volume of space
rapidly, it also gave us uncertainties of up to $\sim$7\arcmin\ in the 
declination
position because the sources could be in the main beam or first sidelobe (or
rarely farther out if the source was very bright at 21 cm).
The lack of declination information meant that the sources needed to be
reobserved to determine their positions and fluxes. We subsequently
followed-up all of the candidate sources at the VLA and at Arecibo.

The VLA follow-up consisted of 10 minute snapshots in  D-array made in January
1998 for 99 sources (90 were detected) most with velocities less than 3000 km
s$^{-1}$. The rest of the galaxies in the  detection survey were followed up at
Arecibo. To improve the coordinates  and fluxes from those determined in the
detection survey we observed in  ``DECscan" mode. We chose to track the
telescope at the detection right ascension and drive the telescope
$\pm$7.5\arcmin\ from the driftscan  declination. The telescope was driven in
declination for a total of 3  minutes and spectra were dumped every 10 seconds
providing 18 spectra  separated by  50\farcs\ The resulting RMS sensitivity is
1.5 mJy. Figure  1 shows an example of a follow-up VLA map and a ``DECscan"
observation. 

A complete discussion of the detection and follow-up surveys is presented by 
Rosenberg \& Schneider (2000).

\begin{figure}
\plottwo{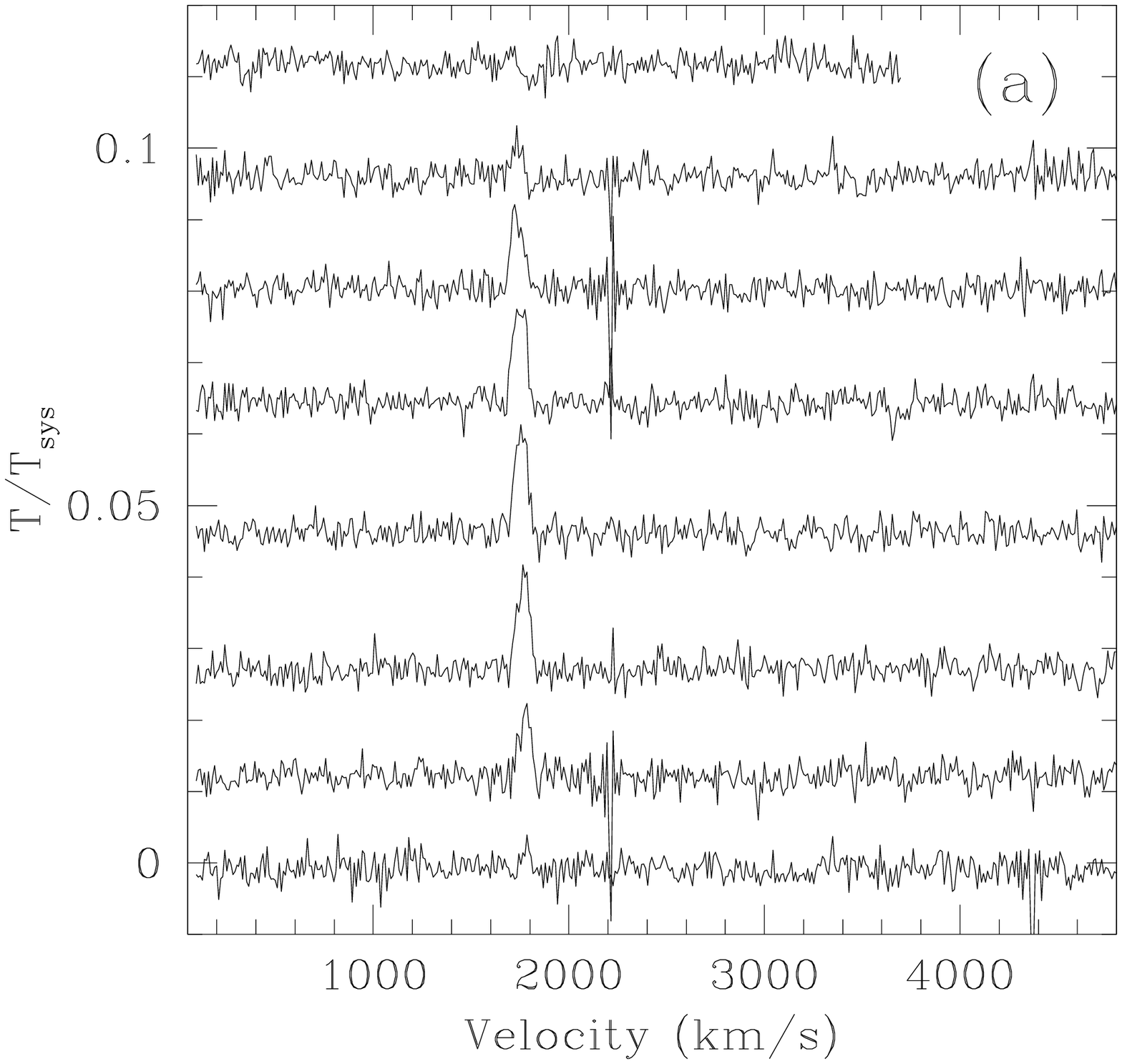}{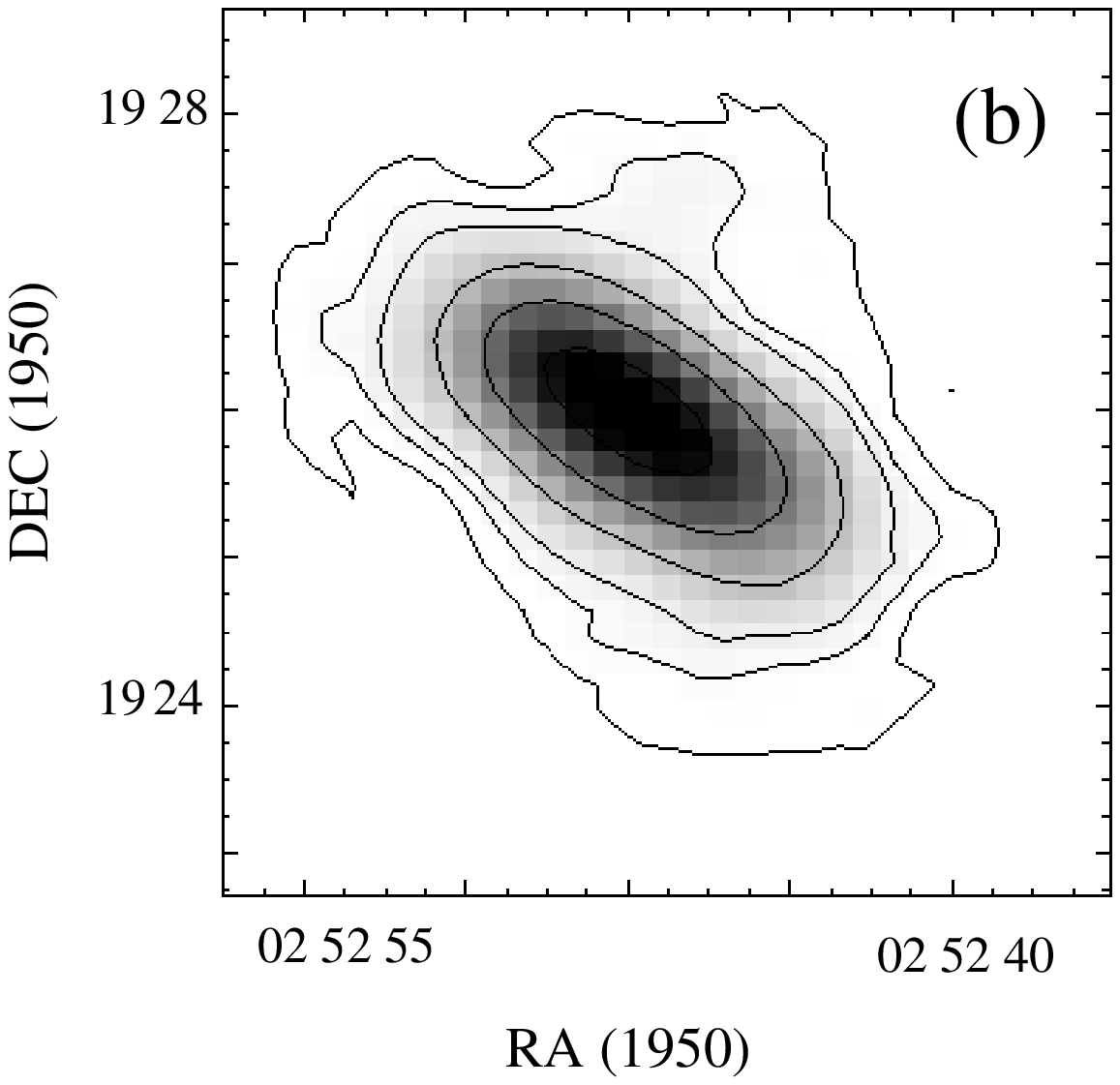}
\caption{(a) An Arecibo ``DECscan" observation. For the Arecibo follow-ups of
the ADBS detections the telescope was fixed at the detection RA and driven in
declination with spectra dumped every 10 seconds. (b) An example of a VLA 
D-array follow-up of the ADBS detections.}
\end{figure}

\section{Synthetic Sources}

We have made use of synthetic sources as a means of analyzing our sample
completeness which is vital for determining the mass function of galaxies.
The sources that were used covered a wide range of luminosities and line 
widths and were given realistic line profiles. The sources were then inserted 
into the data prior to the reduction procedure to be detected along with the
real sources. Figure 2 shows the resulting completeness statistics for 
these sources. The sample is not complete even at high signal-to-noise 
levels because sources at the edges of the bandpass were sometimes fit as 
part of the continuum level. Also, the completeness does not have a sharp 
edge, but falls off fairly rapidly for sources at $<$ 8$\sigma$. The shape
of this curve is very important for the determination of the mass function
(Schneider \& Rosenberg, this volume).

\begin{figure}
\plotfiddle{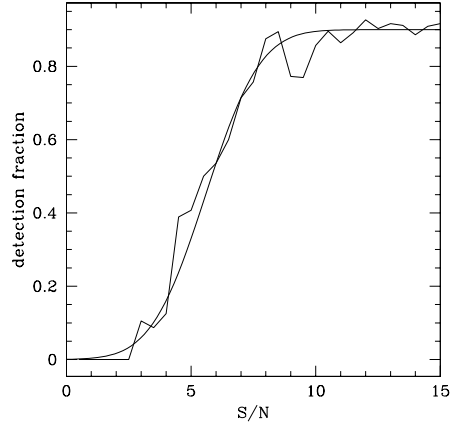}{2.3in}{0}{35}{35}{-110}{-60}
\caption{The relationship between \% completeness and S/N determined from the
synthetic sources that were inserted into the data before the data processing.
The dashed line is a smoothed version of the data and the solid line is a fit.
Note that the completeness never reaches 100\% and falls off precipitously 
below 8$\sigma$.}
\end{figure}

\section{HI-Dominant Galaxies}

We define HI-dominant galaxies as those with ratios of M$_I$/M$_{HI} <$ 1.
Spitzak \& Schneider (1998) studied the properties of these galaxies in their
HI-selected sample and found that they tend to be similar to, although more
extreme than, optically-selected LSBs. As HI-dominance increases, surface
brightness tends to diminish and the galaxies' colors tend to be bluer. Given
the apparently extreme nature (relative to optical samples) of the HI-selected
Spitzak \& Schneider (1998) sample, we would like to use the larger ADBS survey
to address the question: Do HI surveys detect a population of galaxies that is
different from what is found in optical surveys? 

\begin{figure}
\plottwo{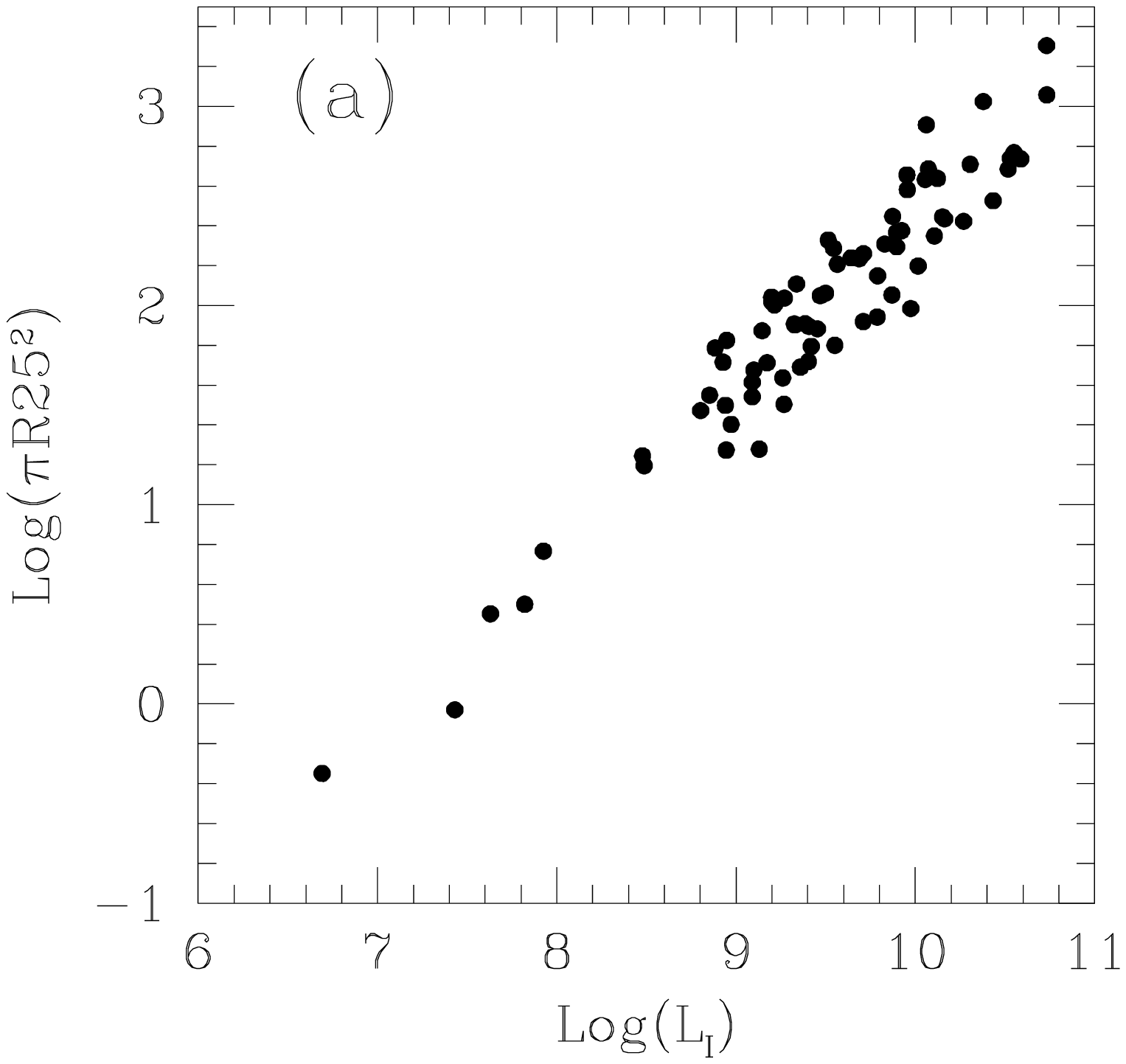}{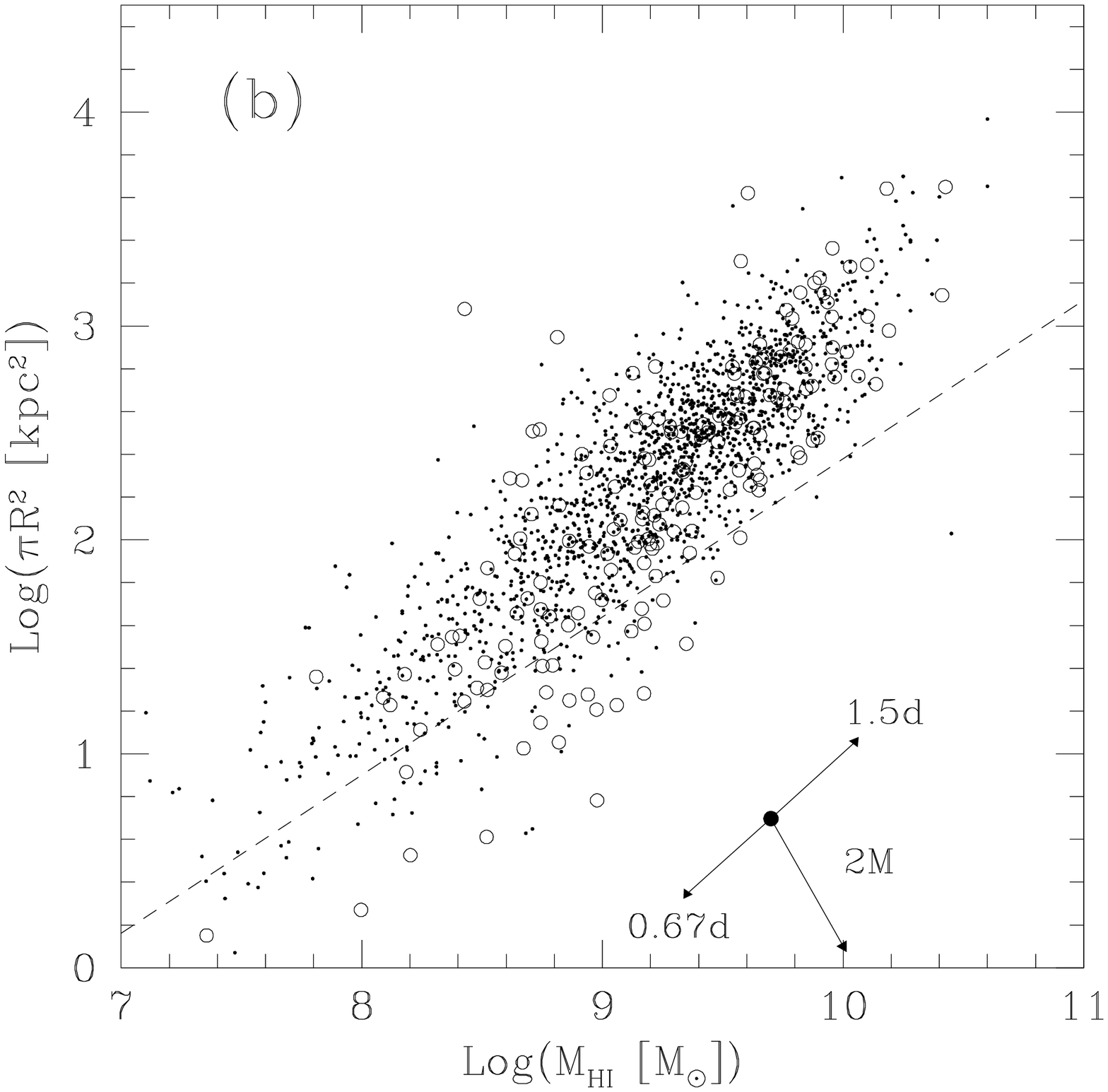}
\caption{(a) The relationship between I-band luminosity and optical area
determined from the Spitzak \& Schneider sample. (b) The relationship 
between HI mass and optical area. The dots are LSB and dwarf galaxies, the open
circles are the ADBS sources. The dashed line shows L$_I$/M$_{HI}$ = 1 assuming
a linear fit to Figure 4a. The errorbars show the effect of  systematic errors
in distance and mass on the data.} \end{figure}

Since we do not currently have optical photometry for these galaxies, we need
to find a surrogate for I-band luminosity. What we have for the ADBS galaxies is
size measured on the Digitized Sky Survey (DSS) images. Figure 3a shows that
there is a tight correlation between log(L$_I$) and  log(area) for the galaxies
in the Spitzak \& Schneider sample. Figure 3b shows the relationship between
optical area and HI mass for optically selected LSB and dwarf galaxies from
Schneider et al. (1992, 1990; filled dots) and ADBS galaxies (open circles). The
dashed line in Figure 3b is approximately L$_I$/M$_{HI}$ = 1 as derived from the
Spitzak \& Schneider relationship. If the LSB and dwarf galaxy sample and the
ADBS select the same populations of HI-dominant galaxies, then there should be
no statistically significant difference in the percentage of galaxies  below and
above the L$_I$/M$_{HI}$ = 1 at a given mass. What we find instead is that
$\sim$27\% of the galaxies in the ADBS sample are below the line in the mass
range between log(M$_{HI}$) = 8.8 and log(M$_{HI}$)  = 9.2 while only $\sim$5\%
of the LSB and dwarf sources are below the line. Since we know that the ADBS is
nearly complete in the 10$^9$ M$_{\sun}$ range, the lack is due to an optical
bias that results in incompleteness even in this LSB and dwarf sample. We expect
this incompleteness to increase at lower HI masses (Schneider \& Schombert
2000). 

\begin{figure}
\plotfiddle{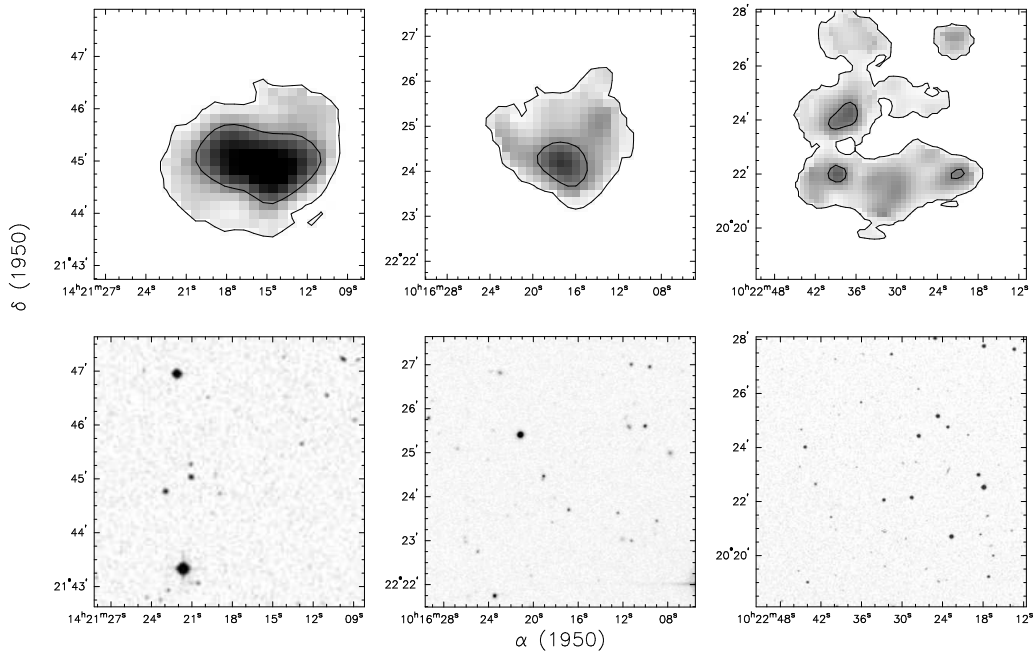}{3.4in}{0}{55}{55}{-520}{-30}
\caption{VLA D-array maps (top row) and the corresponding DSS images 
(bottom row) for 3 HI-dominant galaxies in the ADBS survey.}
\end{figure}

Figure 4  shows some examples of the HI-dominant galaxies that we find in the
ADBS. The top panels are the VLA D-array maps for these galaxies while the lower
panels are the corresponding DSS images. For these galaxies there is little or
no optical emission obviously associated with the HI source. In the end, there
appear to be very few, if any HI sources with no optical counterparts in deep
optical images, but many of the ``possible"  optical counterparts are extremely
small and faint. One of the main  reasons for the incompleteness even of LSB
samples is the minimum angular size requirement. Even if it is possible to
detect these HI sources optically they are very difficult to detect and there
is substantial incompleteness in the selected population.

It is important to detect and study the HI-dominant galaxy population in a
complete and unbiased way because of the significance of these galaxies 
for understanding galaxy populations and evolution. Galaxy evolution is about
the conversion of reservoirs of HI gas to stars. Figure 5 shows the relationship
between the gas-to-total-mass and stars-to-total-mass ratios. This figure 
shows that there is a fairly consistent ratio of baryonic mass to total mass.
The most extreme galaxies on either end of this plot are missing because the
most HI-dominant galaxies are not measured optically while the most
stellar-dominated galaxies do not have detectable HI. In order to have a 
complete understanding of the evolutionary relationship between gas and 
stars we must probe both ends of this distribution. The best way to probe 
the HI-dominant end in an unbiased manner is to use HI-selected samples.

\begin{figure}
\plotfiddle{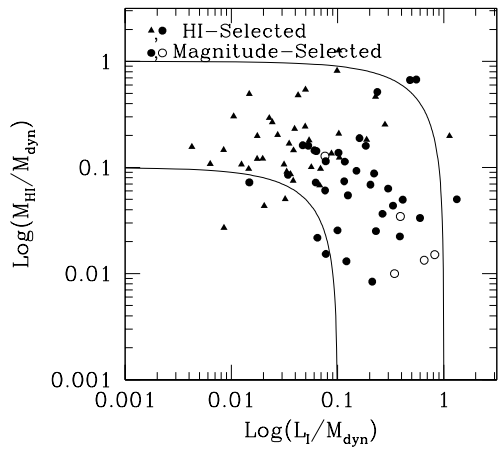}{2.3in}{0pt}{40}{40}{-130}{-90}
\caption{The relation between stellar mass, HI mass, and total mass in
galaxies. The solid lines show M$_{HI}$+M$_I$ equals 0.1 M$_{tot}$ and 
1.0 M$_{tot}$.}
\end{figure}

\section{Conclusions}

We have completed a survey of $\sim$ 430 deg$^2$ at 21 cm and have detected 
265 galaxies, 7 of which are below 10$^8$ M\sun. In order to understand the
significance of these low mass sources and to generally understand the 
statistics of our detection survey, we have used synthetic sources. These
sources allow us to map out our completeness as a function of signal-to-noise.

We have also compared the detection of HI-dominant galaxies in the ADBS to
their detection in the Schenider et al. (1992, 1990) dwarf and LSB sample. We
find that when we select galaxies by their HI, a method that is completely
unbiased by the galaxy's surface brightness, we find a much larger fraction of
the galaxy population to be HI-dominant than is found by the LSB and dwarf
galaxy sample. In addition, the ADBS galaxies are more extreme in their
HI-dominance, surface brightness, and color characteristics than those  detected
in the LSB and dwarf galaxy sample. The LSB and dwarf galaxy survey, and
probably other optical LSB surveys, is not finding all of the galaxies and it is
not detecting them in a manner  which is unbiased by their surface brightness
properties. In order to  understand the processes of galaxy formation and
evolution we must understand the range of star forming environments for galaxies
including the HI-dominant galaxies found in ``blind" HI surveys.

\end{document}